\documentclass[prl,superscriptaddress,twocolumn,showpacs]{revtex4}
\usepackage{epsfig}
\usepackage{bm}
\usepackage{amsmath}

\begin{document}

\setcounter{MaxMatrixCols}{10}

\newcommand{\veck}{\mathbf{k}}
\newcommand{\vecA}{\mathbf{A}}
\newcommand{\vecp}{\mathbf{p}}
\newcommand{\vecr}{\mathbf{r}}
\newcommand{\half}{\frac{1}{2}}
\newcommand{\Eret}{E_\mathrm{ret}}
\newcommand{\Ebind}{E_0^{(N)}}
\newcommand{\vecP}{\mathbf{P}}
\newcommand{\tth}{\Delta t_\mathrm{th}}
\newcommand{\tcm}{\Delta t_\mathrm{rec}}
\newcommand{\ttot}{\Delta t_\mathrm{tot}}

\title{Attosecond electron thermalization by laser-driven electron recollision in atoms}

\author{X. Liu}
\affiliation{Department of Physics, Texas A\&M University, College
Station, TX 77843-4242, USA}
\author{C. Figueira de Morisson Faria}
\affiliation{Centre for Mathematical Science, City University,
Northampton Square, London EC1V OHB, UK}
\author{W. Becker}
\affiliation{Max-Born-Institut,
Max-Born-Str. 2A, 12489 Berlin, Germany}
\author{P. B. Corkum}
\affiliation{National Research Council of Canada, 100 Sussex Drive,Ottawa, Ontario, Canada K1AOR6}
\date{\today}

\begin{abstract}
Nonsequential multiple ionization of atoms in intense laser fields is initiated by a recollision between an electron, freed by tunneling, and its parent ion.  Following recollision, the initial electron shares its energy with several bound electrons.  We use a  classical model based on rapid electron thermalization to interpret recent experiments.  For neon, good agreement with the available data is obtained with an upper bound of 460 attoseconds for the thermalization time. 
\end{abstract}
\maketitle

Atoms exposed to intense laser fields ionize. The freed electron and its ionic partner are
accelerated by the laser field away from each other. When the field
changes  sign, it may drive the electron into a recollision with the
ion. This simple mechanism, which for high-intensity low-frequency
fields is largely classical, governs many laser-atom processes such
as high-order harmonic generation (HHG), high-order above-threshold
ionization (HATI), and nonsequential double ionization (NSDI) and
explains the gross features of the spectra observed \cite{corkum}.

The period of the commonly applied titanium-sapphire (Ti:Sa) laser
is about 2.7 fs. The recollision physics unfold on the
time scale of a small fraction of the laser period. Therefore, the analysis
of laser-induced recollision phenomena provides access to the
inner-atomic dynamics on the attosecond time scale and, indeed,  the focus of
recent investigations has moved into this temporal domain. For
example, it has brought molecular imaging with subangstrom spatial and
subfemtosecond temporal resolution within reach \cite{Niikura,molimag}. The
advent of phase-stabilized infrared few-cycle pulses and of uv pulses of attosecond
duration allows even more control \cite{Baltuska}, but neither are
necessary for a study of the attosecond dynamics.

In this Letter, we analyze recent experiments on nonsequential multiple ionization (NSMI) of neon [5,6] in which the momentum distributions of Ne$^{N+}$ $(N=3,4)$  were measured at two intensities.  We use the fact that the time-dependent laser field provides a clock --- the field accelerates the ion to a final velocity that depends on the times at which the $N$ electrons ionized.  This ``streak camera'' \cite{streak} therefore measures the range of times of ionization.  Comparing the measured momentum distributions to those predicted by a classical model, we infer that the recolliding electron thermalizes with the $N-1$ bound electrons in less than 500 attoseconds.  To our knowledge, no other method allows such a low upper bound for thermalization times within atoms to be measured.

Nonsequential double and multiple ionization is defined by the fact
that it is not sequential, that is, it is not the product of a
sequence of uncorrelated single-ionization events. NSDI and NSMI
require electron-electron correlation as a necessary precondition \cite{Ffm2000,FrMBI2000}.  
Even for the very simplest such process -- NSDI of helium -- a fully
quantum-mechanical description from first principles, i.e. by
solution of the time-dependent Schr\"odinger equation in six spatial
dimensions,  has not been accomplished yet \cite{taylor}, and for
the heavier atoms it is clearly out of the question. This leaves
approximate quantum-mechanical approaches, such as density-functional methods
\cite{bauer}, $S$-matrix methods that try to identify the most
relevant terms of an appropriate perturbative expansion
\cite{faisal,KBRS}, or classical-trajectory methods
\cite{china,eberly}. For \textit{multiple}
(triple and higher) ionization, which  occurs under the same
conditions as NSDI if the laser intensity is high enough,
any description from first principles appears to be utterly out of
reach. It seems equally hopeless, for three electrons and more, to
identify the relevant diagrams in the microscopic $S$-matrix
approach.

 The model we here propose is in the spirit
of the one of Ref.~\cite{KBRS}, but essentially classical. 
We investigate the scenario
wherein  NSMI is effected by one single recollision. 
We assume that the pertinent
electron tunnels into the continuum with zero velocity at the ionization time $t'$ according to  
the time-dependent rate $R(t')$, for which we adopt the standard quasi-static rate \cite{LL}. Thereafter, we turn to an entirely classical description: 
 The laser field may drive the
electron back to its parent ion at a later time $t$, which is a function $t(t')$ of the ionization time and can be easily evaluated. 
We assume that the energy $E_\mathrm{ret}(t)$ of the returning
electron be completely thermalized among the ensemble of 
participating electrons, that is, the returning electron and the
$N-1$ electrons to be freed. These $N$ electrons then form an excited complex with
the total energy (with respect to the continuum threshold) $E_\mathrm{ret}(t) - E_0^{(N)}$, where $E_0^{(N)}>0$ denotes the total ionization potential of the $N-1$ (up to the recollision time $t$ inactive) electrons. The distribution of energy and momentum
over the $N$ electrons is assumed to be completely statistical and only
governed by the available phase space. At the time $t+\Delta t$, the $N$ electrons become free to move in the laser field, which is described by the vector
potential $\vecA(t)$ such that $\vecA(t)=\mathbf{0}$ outside the pulse.

The corresponding distribution of the final electron momenta 
$\vecp_n\ (n=1,\dots,N)$ is proportional to 
\begin{eqnarray}
F(\vecp_1,\vecp_2,\dots,\vecp_N) = \int dt' R(t') \delta \left(E_0^{(N)} -E_{\rm ret}(t)\right. \nonumber\\
\left.+\half \sum_{n=1}^N [\vecp_n+\vecA(t+\Delta t)]^2 \right), \label{F}
\end{eqnarray}
where the integral extends over the ionization time $t'$. The $\delta$ function expresses the fact that the total kinetic energy of the $N$ participating electrons at the time $t+\Delta t$ is fixed by the first-ionized electron at its recollision time $t$. This constitutes the one and only condition on the final momenta $\vecp_n$. The only free parameter of this model is the time delay $\Delta t$ between the recollision time and the time when the electrons become free. It is the sum of a thermalization time $\tth$  -- the time it takes to establish the statistical ensemble -- and a possible additional ``dwell time'', until  the electrons become free. By comparing the predictions of the model with the data, we will be able to infer a value of $\Delta t$, which in turn provides an upper limit for the thermalization time $\tth$.

This model is an extension to NSMI of a classical model introduced for NSDI   for $\Delta t=0$ \cite{FFetal04R,FFetal04}.  Sufficiently high
above threshold, it produced momentum distributions that were virtually
indistinguishable from their quantum-mechanical counterparts.
Statistical models similar to the one above have been used in many areas of 
physics. For example, the statistical Rice-Ramsperger-Kassel-Marcus (RRKM) theory \cite{RRKM} describes thermalization of molecular vibrational degrees of freedom, and for  high-energy collisions of elementary particles
 and heavy ions statistical models have been utilized to  predict the momentum spectra
of the reaction products \cite{hagedorn}. An excited complex as the
doorway to NSMI was also  considered in Ref.~\cite{sachaeckh}.

 A convenient feature of the ansatz (\ref{F}) is
that integration over unobserved momentum components is 
easily carried out. To this
end, we exponentialize the $\delta$ function in Eq.~(\ref{F}) with
the help of its Fourier representation
\begin{equation}
\delta (x) = \int_{-\infty}^\infty \frac{d\lambda}{2\pi} \exp (-i\lambda x).\label{delta}
\end{equation}
Infinite integrations over the momenta $\vecp_n$ can then be done 
by Gaussian quadrature. The remaining integration over the
variable $\lambda$ is taken care of by the formula \cite{GR}
\begin{equation}
\int^\infty_{-\infty}\frac{d\lambda}{(i\lambda+\epsilon)^{\nu}} e^{ip\lambda}=\frac{2\pi}{\Gamma (\nu)}p^{\nu -1}_+,
\end{equation}
where $x^\nu_+ = x^\nu \theta(x)$, with $\theta(x)$ the unit step
function  and $\epsilon \rightarrow +0$.

For comparison with the experiments \cite{FrMBI2000,multi}, we
calculate the distribution of the momentum $\vecP$ of the ion.
Provided the momentum of the absorbed laser photons can be
neglected, momentum conservation implies $\vecP=-\sum_{n=1}^N
\vecp_n$, so that
\begin{eqnarray}
F_\mathrm{ion}(\vecP) \equiv \int \prod_{n=1}^N d^3\vecp_n \delta
\left(\vecP+\sum_{n=1}^N \vecp_n \right) F(\vecp_1,\vecp_2,\dots,
\vecp_N)\nonumber\\
=\frac{(2\pi)^{\frac{3N}{2}-\frac{3}{2}}}{N^{3/2}\Gamma(\frac32(N-1))}\int
dt' R(t') \left(\Delta
E_{N,\mathrm{ion}}\right)^{\frac{3N}{2}-\frac{5}{2}}_+\label{ion}
\end{eqnarray}
with $\Delta E_{N,\mathrm{ion}} \equiv E_\mathrm{ret}(t) -\Ebind-\frac{1}{2N}[\vecP-N\vecA(t+\Delta t)]^2$.

In the experiments thus far, the ion momentum transverse to the
direction of the laser polarization is entirely or partly integrated
over. In the first case, the remaining distribution of the
longitudinal ion momentum $P_\parallel$ is
\begin{eqnarray}
F_\mathrm{ion}(P_\parallel) \equiv \int d^2\vecP_\perp F_\mathrm{ion}(\vecP)\nonumber\\
=\frac{(2\pi)^{\frac{3N}{2}-\frac{1}{2}}}{\sqrt{N}
\Gamma((3N-1)/2)}\int dt' R(t') \left(\Delta
E_{\mathrm{ion}\parallel}\right)^{\frac{3N}{2}
-\frac{3}{2}}_+,\label{ionpar}
\end{eqnarray}
where now $\Delta E_{\mathrm{ion}\parallel} \equiv E_\mathrm{ret}(t)
-\Ebind-\frac{1}{2N}[P_\parallel -NA(t+\Delta t)]^2$. If just one
transverse-momentum component ($P_{\perp,2}$, say) is integrated
while the other one ($P_{\perp,1} \equiv P_\perp$) is observed, the
corresponding distribution is
\begin{eqnarray}
F_\mathrm{ion}(P_\parallel,P_\perp) \equiv \int dP_{\perp,2} F_\mathrm{ion}(\vecP)\nonumber\\
=\frac{(2\pi)^{\frac{3N}{2} -1}}{N\Gamma(3N/2-1)} \int dt' R(t')
\left(\Delta E_{N,\mathrm{ion}\parallel\perp}
\right)^{\frac{3N}{2}-2}_+\label{ionparperp}
\end{eqnarray}
with $\Delta E_{N,\mathrm{ion}\parallel\perp} \equiv
\Delta E_{\mathrm{ion}\parallel}-\frac{1}{2N}P_\perp^2$.

\begin{figure}
\includegraphics[width=7cm, clip=true]{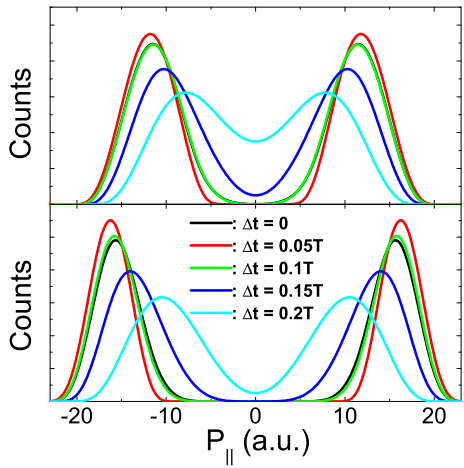}
\caption{(Color online) Distribution of the longitudinal ion momentum for triple (upper panel) and quadruple (lower panel)  nonsequential ionization of neon 
at 2 PWcm$^{-2}$ calculated from Eq.~(\ref{ionpar}) for various delays $\Delta t$ as indicated in the lower panel. Note that in the upper panel the curves for $\Delta t=0$ and $\Delta t=0.1T$ almost completely overlap.}
\label{fig1}
\end{figure}

\begin{figure}
\includegraphics[bb=00 15 300 280, clip=true, width=7.5cm]{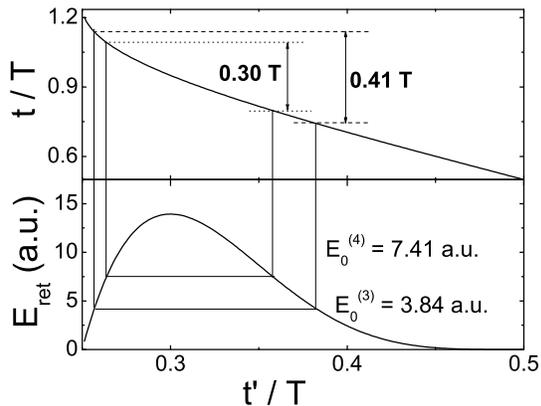}
\caption{The kinetic energy $E_\mathrm{ret}(t)$ (lower panel) and recollision time
$t$ (upper panel) of electrons liberated at the tunneling time $t'$. The
classically allowed intervals $\tcm$ of the recollision time, for which the
corresponding electron return energy $E_\mathrm{ret}(t)$ is greater than $E_0^{(3)}$ (and $E_0^{(4)}$), are shown in the upper panel. The calculation is for neon at
2.0\,PWcm$^{-2}$.} \label{therm-time}
\end{figure}

In Fig.~\ref{fig1} we present calculations of the ion-momentum 
distributions for 
triple (upper panel) and quadruple (lower panel) NSMI of neon 
according to Eq.~(\ref{ionpar}), for various values of $\Delta t$ 
between 0 and $0.2T$. The parameters
are for the experimental data of neon presented in
Fig.~2 of Ref.~\cite{multi}. The figure displays the  
double-hump structure of the ion momentum in NSMI of neon. When 
the time delay $\Delta t$ increases from 0, initially, the  effect on the momentum distribution is small. Later, however, the center positions of the humps start moving towards zero momentum and the widths of the humps increase until the two humps begin to merge. This behavior can easily be understood from the recollision kinematics, which are illustrated in Fig.~\ref{therm-time}. The final electrons undergo maximal acceleration if they are released near a zero crossing of the electric field, which in the figure occurs at $t=T$. For the example of triple ionization of neon, the earliest recollision 
with $E_\mathrm{ret}> E_0^{(3)}$ takes place at $t=0.74T$. Already with a delay of $\Delta t > 0.26T$, all electrons will be released after the zero crossing, which results in significantly lower ion momenta.

\begin{figure}
\includegraphics[clip=true, width=6.5cm]{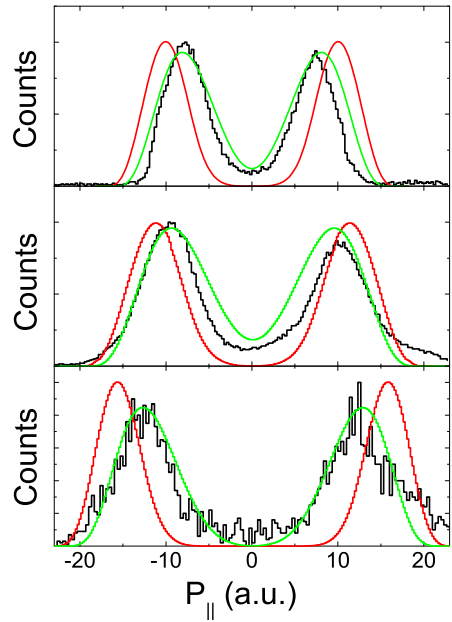}
\caption{(Color online) Distribution of the longitudinal momentum of nonsequential
triple ionization of neon at 1.5 PWcm$^{-2}$ (upper panel),
2.0 PWcm$^{-2}$ (middle panel), and of nonsequential quadruple
ionization of Ne at 2.0 PWcm$^{-2}$ (lower panel). The rugged (black) curves
represent the data of Fig. 2 of Ref.~\cite{multi}. The outermost smooth (red) curve and the innermost smooth (green) curve are calculated from Eq.~(\ref{ionpar}) for $\Delta t=0$ and $\Delta t=0.17T$. }
\label{fig3}
\end{figure}

\begin{figure}
\includegraphics[width=6.5cm,clip=true]{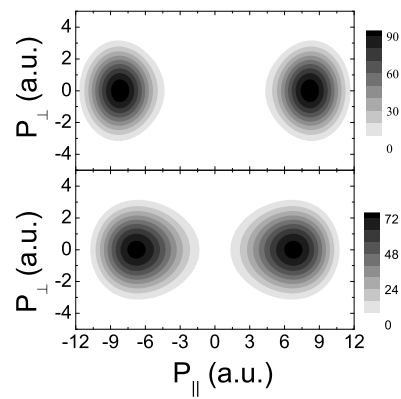}
\caption{Ion-momentum distribution of nonsequential triple
ionization of neon at 1.0 PWcm$^{-2}$ for $\Delta t=0$ (upper panel) and
$\Delta t=0.17T$ (lower panel), calculated  from
Eq.~(\ref{ionparperp}), to be compared with the data of
Ref.~\cite{FrMBI2000}.} \label{neon00}
\end{figure}

In Fig.~\ref{fig3} we compare the results of the statistical model with the data of Ref.~\cite{multi}. The intensities are those given for the experiment. We display momentum distributions calculated from Eq.~(\ref{ionpar}) for zero delay and for $\Delta t= 0.17T$. The latter value was chosen to yield optimal agreement for the entire set of data.  We notice, in particular, that with this nonzero delay the model reproduces the maxima of the experimental ion-momentum distribution. This removes a longstanding discrepancy between models of the type discussed in Refs.~\cite{FFetal04R,FFetal04,KBRS} and the data.  For triple ionization, the calculated momentum distributions are wider than those of the data, in particular for the higher intensity, even though the data may include a contribution of the partly sequential channel Ne $\to$ Ne$^+ \to$
Ne$^{3+}$, or the contribution 
of a recollision-excitation channel, both of which are, of course, not part of the model. 

Figure \ref{neon00} exhibits the results of our thermalization model
for $N=3$ for the conditions of Ref.~\cite{FrMBI2000}.  The
distribution of two components of the ion momentum is presented, the
component parallel to the laser field and one transverse component, which therefore provides a more stringent test of the model. 
The third component is integrated over in the data, which  corresponds to
Eq.~(\ref{ionparperp}). The intensity given in the experiment is
1.5\,PWcm$^{-2}$; we obtain good agreement with the data for the lower intensity of 1.0\,PWcm$^{-2}$ and $\Delta t=0.17T$ as before \cite{footnote}. The nonzero delay $\Delta t$ has little effect on the transverse width of the distribution, but it causes an elongation in the longitudinal direction and moves the centers to lower momenta, markedly improving the agreement with the data. 

Encouraged by the good  agreement between the model and the data, we interpret the delay for which we observed optimal agreement as an upper bound of the
thermalization time $\tth$, as argued above. This yields $\tth < \Delta t_\mathrm{opt} = 0.17T \simeq  460$ as. A lower limit of the thermalization time should be given by the inverse of the plasma frequency for an 
electron density $\rho_\mathrm{e} = N$ in atomic units. This produces $\tth < \sqrt{\pi/N}$, which is of the order of the atomic unit of time.

All of the above discussion has been for neon. NSDI of argon appears to be governed by a recollision-excitation
scenario \cite{ArvsNe}. We note in passing that for (quadruple) NSMI of argon we get good agreement of our statistical thermalization model with the data \cite{multi} for the experimental intensity and $\Delta t=0.35T$, twice as long as for neon. Details will be given elsewhere.

To test the model further, and possibly to set a tighter upper limit on the thermalization time, it is necessary to restrict the time range $\Delta t_\mathrm{CM}$ of recollision  (Fig.~\ref{therm-time}).  If thermalization is more rapid than 460 as, then the width of the ion-momentum distribution will decrease as we restrict $\tcm$.  There are a number of ways to minimize $\tcm$.  Within limits, all that is needed is to lower the light intensity or to increase the laser frequency.  However, the best way to minimize $\tcm$ and to control the time of recollision is to use a second harmonic field, polarized perpendicular to the fundamental.  The combined requirement that the electron and ion recollide in both directions allows the time of recollision to be precisely determined and controlled via the relative phase of the two beams.  

The limit on the thermalization time that we determine (as well as the much tighter bounds that seem feasible in future experiments) should apply to stationary electron-atom scattering in general.  From a collision-physics perspective, as a result of streaking, the time-dependent laser field reveals information that is hard to obtain by other means.  The streaking principle should also be applicable for studying nuclear dynamics.  As in NSMI, nuclear processes also can be initiated by laser-controlled re-collision \cite{nuclear} (of course, at much higher intensity).  Any nuclear decay process that results in a mass or charge change of the fragments will be streaked by the laser field just as electrons and ions are streaked in our case.

To summarize, by comparison of experimental data for triple and
quadruple nonsequential ionization of neon with a simple statistical 
recollision model where the returning electron thermalizes with a
subset of the bound electrons, we have been able to conclude that
(i) for neon such a model appears to contain the most relevant
physics, and (ii) the time for this thermalization to occur is
extremely fast, well below one femtosecond.

We gratefully acknowledge stimulating  discussions with G.G. Paulus, H. Rottke,  and W. Sandner.


\begin{thebibliography}{99}
\bibitem{footnote} A delay of the same order has also been observed in  classical-trajectory simulations of NSDI of helium \cite{eberly}.
\end{thebibliography}
\end{document}